\documentclass[	
						superscriptaddress,%
						a4paper,
						showkeys,
					]{revtex4}

\usepackage{color}
\usepackage{upgreek}					
\usepackage[dvips]{graphicx}		

\newcommand{\etal}{\textit{et al.}}

\begin{document}

\title{High-pressure phase transitions in BiFeO$_3$: hydrostatic vs. non-hydrostatic conditions}

\author{Mael Guennou}
\affiliation{Laboratoire des Mat\'eriaux et du G\'enie Physique, CNRS, Grenoble Institute of Technology, MINATEC,
3 parvis Louis N\'eel, 38016 Grenoble, France}
\author{Pierre Bouvier}
\affiliation{Laboratoire des Mat\'eriaux et du G\'enie Physique, CNRS, Grenoble Institute of Technology, MINATEC,
3 parvis Louis N\'eel, 38016 Grenoble, France}
\affiliation{European Synchrotron Radiation Facility (ESRF), BP 220, 6 Rue Jules Horowitz, 38043 Grenoble Cedex, France}
\author{Rapha\"el Haumont}
\affiliation{Laboratoire de Physico-Chimie de l'\'Etat Solide, ICMMO, CNRS, Universit\'e Paris XI, 91405 Orsay, France}
\author{Gaston Garbarino}
\affiliation{European Synchrotron Radiation Facility (ESRF), BP 220, 6 Rue Jules Horowitz, 38043 Grenoble Cedex, France}
\author{Jens Kreisel}
\affiliation{Laboratoire des Mat\'eriaux et du G\'enie Physique, CNRS, Grenoble Institute of Technology, MINATEC,
3 parvis Louis N\'eel, 38016 Grenoble, France}

\begin{abstract}
We report high-pressure x-ray diffraction experiments on BiFeO$_3$ (BFO) single crystals in diamond-anvil cells up to 14 GPa. Two data sets are compared, one in hydrostatic conditions, with helium used as pressure-transmitting medium, and the other in non-hydrostatic conditions, with silicon oil as pressure-transmitting medium. It is shown that the crystal undergoes different phase transitions in the two cases, highlighting the high sensitivity of BFO to non-hydrostatic stress. Consequences for the interpretation of high-pressure structural studies are discussed.
\end{abstract}


\maketitle

\section{Introduction}

Bismuth ferrite (BiFeO$_3$ - BFO) exhibits at ambient conditions both ferroelectric and magnetic order and has become the archetype of an intrinsic multiferroic, studied for both technology-driven and fondamental research \cite{Catalan2009}. From a structural point of view, BFO crystallizes at ambient conditions in the perovskite structure with space group $R3c$. This structure can be derived from the cubic parent perovskite by rotations (tilts) of the oxygen octaedra around the $[111]_c$ direction relative to the parent cubic cell ($a^-a^-a^-$ in Glazer's notations) and displacements of the Bi$^{3+}$ and Fe$^{3+}$ cations along the same $[111]_c$ direction. 

Here we are interested in the effect of high-pressure, a parameter which has been much less explored in the past, when compared to the wealth of temperature-dependent measurements, and this although Samara's pioneer work \cite{Samara1975,Samara1996,Samara2000} has amply illustrated the usefulness of this parameter. More recently, the parameter high-pressure has attracted more interest \cite{Kreisel2009}, as exemplified by studies on classic ferroelectrics \cite{Kornev2005,Janolin2008,Ahart2008,Ravy2007}, ferroelastics \cite{Angel2005,Bouvier2002,Guennou2010,Guennou2010a}, piezoelectrics \cite{Sani2004,Rouquette2005,Rouquette2004,Fraysse2008} etc. Although these recent studies have led to a better understanding of the effect of high-pressure on cation displacements in ferroelectrics or octahedra tilts in ferroelastics, there is only little work on materials that present simultaneously those two competing structural instabilities. BFO is one of the very few perovskites that exhibits both cation displacements and octaedra tilts at ambient conditions, making it an ideal candidate for this field of study. One of the consequences of this complex competition in BFO is a rich pressure-temperature phase diagram which is still not completely understood ; not only the high-temperature regime but also the effects of a high hydrostatic pressure remain controversial \cite{Catalan2009}. In this study, we will focus on the effect of low and medium pressures (up to 14 GPa) at room temperature, aiming at a better understanding of the literature discrepancies, which are summarized in the following.

Powder x-ray diffraction under high-pressure has been performed over the past few years by several groups with various outcomes. Gavriliuk \etal{} \cite{Gavriliuk2007,Gavriliuk2008} and Zhu \etal{} \cite{Zhu2010} have reported no phase transition in this pressure range. Haumont \etal{} \cite{Haumont2009} have identified two phase transitions at 3.5 and 10 GPa with the phase sequence $R3c\longrightarrow C2/m\longrightarrow Pnma$. Belik \etal{} \cite{Belik2009} have later confirmed the phase transition at 4 GPa and identified an additional transition at 7 GPa. They have identified the two intermediate phases as orthorhombic instead of the monoclinic $C2/m$, with a transition sequence $R3c\longrightarrow\mathrm{Ortho\ I}\longrightarrow\mathrm{Ortho\ II}\longrightarrow Pnma$, and also pointed out reversibility issues. In parallel, Raman spectroscopy has also been used to investigate phase transitions \cite{Haumont2006,Yang2009}. Both studies revealed two phase transitions in the low-pressure range, but at somewhat different pressures, about 3 and 9 GPa. 

This variety of results reported in the literature suggests that the structure of BFO with its competing instabilities is particularly sensitive to the quality of the sample and experimental conditions (stoichiometry, homogeneity and isotropy of the stress field etc.). In order to evaluate the effect of non-hydrostatic pressure conditions as one source of the controversy, we compare in this work two high-pressure single-crystal x-ray diffraction experiments carried out in diamond anvil cells with different pressure-transmitting media. This comparison suggests that the quality of the hydrostatic conditions is one source explaining the variety of results in the literature on BFO.

\section{Experimental aspects}

BFO single crystals were grown by the flux method as described elsewhere \cite{Haumont2008}. The samples used for the experiments were polished to a thickness of about 10 $\upmu$m with a lateral extension of 10 to 30 $\upmu$m. No domain structure was visible before the experiments, neither under polarized light nor in the initial diffraction patterns. All experiments were performed in diamond-anvil cells (DAC). The diamonds have the Boehler-Almax design with a cullet of 600 $\upmu$m. The pressure chamber was sealed by a stainless steel gasket pre-indented to a thickness of about 60 $\upmu$m. 

X-ray diffraction experiments were performed on the ID27 and ID09A beamlines at the ESRF. At ID27, the beam was monochromatic with a wavelength of 0.3738 \AA\ selected by an iodine K-edge filter and focused to a beam size of about 3 $\upmu$m. At ID09A, the beam size is about 20 $\upmu$m and the wavelength (0.4144 \AA) was determined from the calibration using a standard silicon powder. The signal was collected in the rotating crystal geometry on a CCD detector with $-30^\circ\le\omega\le 30^\circ$ or $-24^\circ\le\omega\le 24^\circ$, depending on the opening of the cell used. A precise calibration of the detector parameters was performed with a reference silicon powder. The diffraction patterns were indexed with a home-made program based on the Fit2D software \cite{Hammersley1996}. The refinement of the lattice constants from the peak positions was performed with the program UnitCell \cite{UnitCell}.

\begin{figure}[htbp]
\begin{center}
\includegraphics[width=0.4\textwidth]{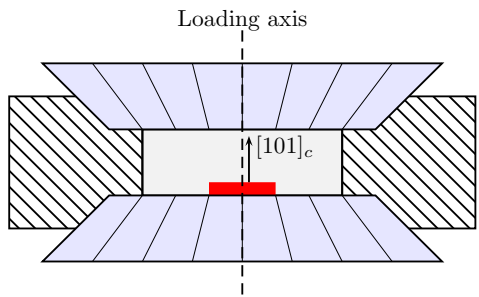}%
\caption{Sketch of the diamond-anvil cell (DAC) with the orientation of the sample relative to the loading axis of the cell. The sample is immersed into a pressure-transmitting medium (PTM): helium and silicon oil for hydrostatic and non-hydrostatic conditions, respectively.}
\label{fig:DAC}
\end{center}
\end{figure}

For a measurement in best hydrostatic conditions, helium was used as a pressure transmitting medium (PTM). Helium remains liquid up to 12.7 GPa. As such, it provides perfect hydrostatic conditions below this limit and remains very good up to very high pressures \cite{Angel2007,Klotz2009}. For the measurement in non-hydrostatic conditions, we used silicon oil as a PTM, which is a very viscous fluid that freezes rapidly. Its hydrostatic limit has been reported to lie around 3 GPa, as determined from the analysis of the ruby luminescence lines \cite{Klotz2009}, or even 0.9 GPa from the measurement of diffraction line broadening induced in a quartz reference crystal \cite{Angel2007}. Above this limit, the stress can no longer be regarded as hydrostatic and a so-called deviatoric component adds to the hydrostatic stress field. With the $z$-axis along the loading axis of the cell, the total stress field $\sigma$ can be written \cite{Zhao2010}:
\[
\sigma =
\left(
\begin{array}{c c c}
-P & &  \\
 & -P & \\
 & & -P \\
\end{array}
\right)
+
\left(
\begin{array}{c c c}
\sigma_D & &  \\
 & \sigma_D & \\
 & & -2\sigma_D \\
\end{array}
\right)
\]
where $P$ and $\sigma_D$ measure the magnitude of the hydrostatic pressure and deviatoric stress respectively. The deviatoric stress has a rotational symmetry around the loading axis of the cell, with a stronger compressive stress along the axis. In an experiment with silicon oil as PTM, Zhao \etal{} \cite{Zhao2010} have estimated this stress to be of 1 GPa at 10 GPa approximately. Although our setup is not strictly identical, this value can be considered as a reasonable estimate for the maximum deviatoric stress in our experiment. 

For a non-hydrostatic stress field, the effect of high pressure is expected to depend on the orientation of the crystal with respect to the loading axis of the DAC. In that case, the loading axis is directed perpendicular to a natural face of the as-grown crystal, close to $[101]_c$. The setup with the orientation of the crystal is sketched in figure \ref{fig:DAC}. 

In all the experiments, the fluorescence of ruby was used as a pressure gauge \cite{Mao1986}. Note that as the non-hydrostatic stress sets on, the pressure measurement from the ruby line becomes slightly inaccurate, but it remains within negligible uncertainties \cite{Zhao2010} and does not alter the conclusions presented in this study. 

\section{Results}

\begin{figure}[htbp]
\begin{center}
\includegraphics{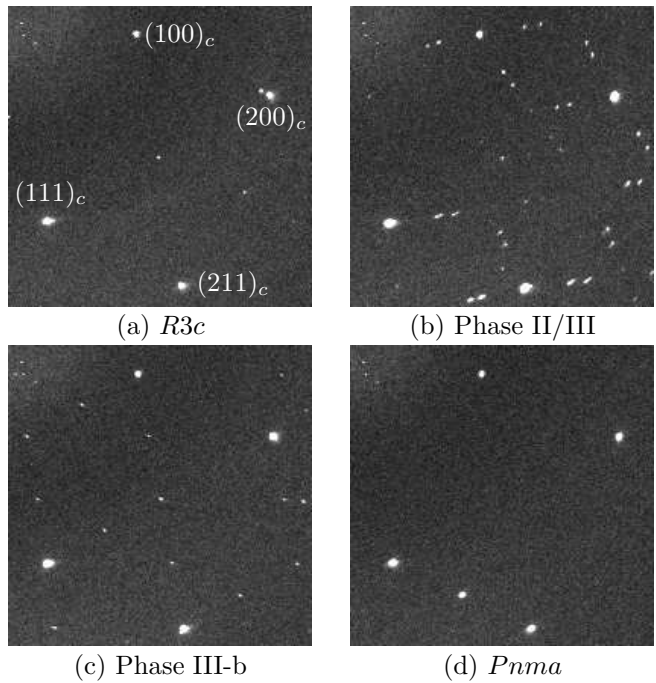}%
\caption{Diffraction patterns observed for the different phases. For clarity, all patterns are taken from the same experiment, namely with non-hydrostatic conditions. Patterns (a), (b) and (d) are qualitatively identical when the conditions are hydrostatic. Pattern (c) was only observed in non-hydrostatic conditions. The indices given in pattern (a) are relative to the parent cubic cell.}
\label{fig:cliches}
\end{center}
\end{figure}

\begin{figure}[htbp]
\begin{center}
\includegraphics[width=0.75\textwidth]{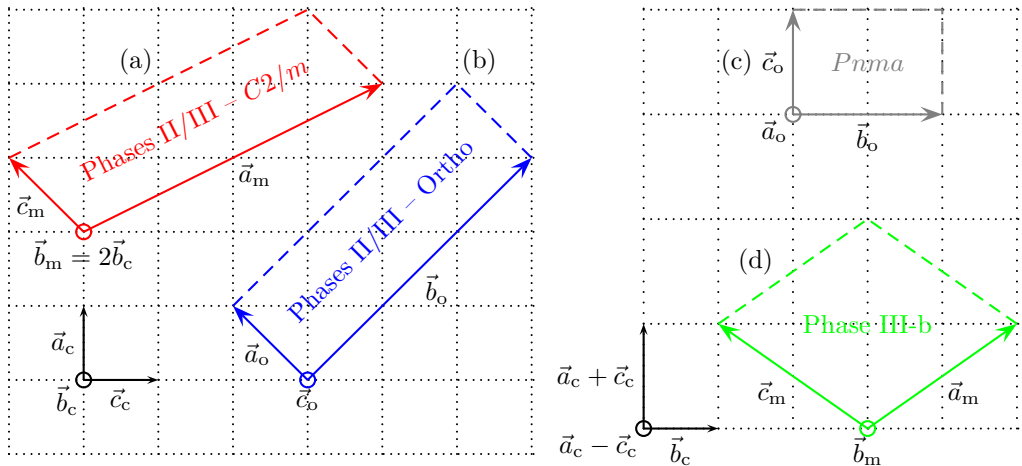}%
\caption{Relations between the parent cubic cell and the unit cells proposed. (a) and (b) Phase II and III. The two cells pictured are the monoclinic $C2/m$ (a) and orthorhombic (b) cells proposed in \cite{Haumont2009} and \cite{Belik2009} respectively. (c) $Pnma$ (d) Phase III-b.}
\label{fig:mailles}
\end{center}
\end{figure}

\begin{figure}[tbp]
\begin{center}
\includegraphics[width=0.45\textwidth]{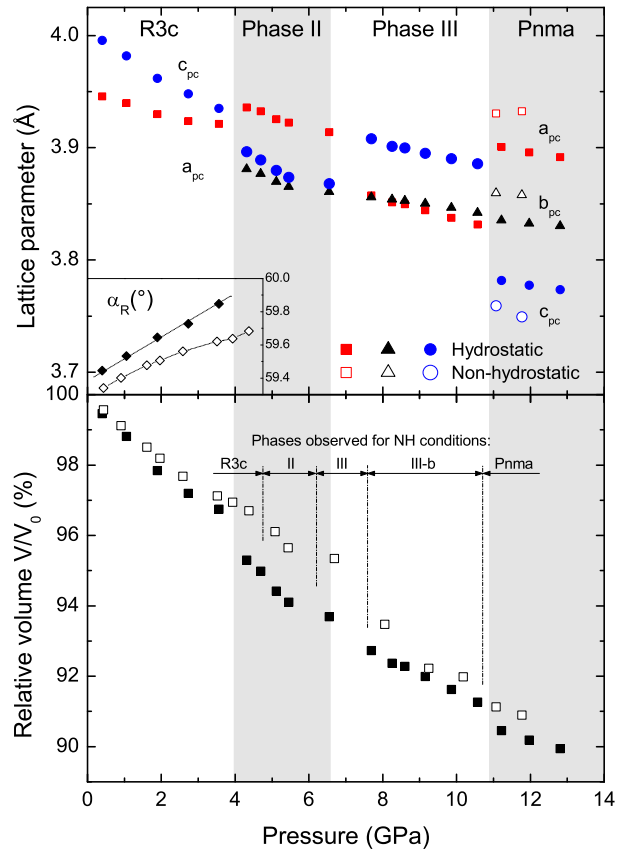}%
\caption{Evolution under pressure of the pseudo-cubic lattice constants (top), rhomboedral angle (inset) and relative volume (bottom). The full and open symbols correspond to hydrostatic and non-hydrostatic conditions respectively. The stability domains of the various phases represented as coloured stripes are given for hydrostatic conditions, while the phase sequence observed in non-hydrostatic conditions is indicated by dashed lines in the volume plot.}
\label{fig:param}
\end{center}
\end{figure}

We start with the description of the phase transitions observed for hydrostatic pressure conditions. At ambient pressure, the crystal is in a single domain state in the familiar rhomboedral structure with space group $R3c$. The diffraction pattern integrated over the full angular range is identical to the pattern shown in figure \ref{fig:cliches} (a). Between room pressure and 4 GPa, the rhomboedral distortion, as measured by the rhomboedral angle $\alpha$, descreases gradually. At 4 GPa, the crystal undergoes a first transition to a new phase which we label phase II. In this phase, the crystal exhibits a more complex diffraction pattern, with the emergence of new superstructure reflections (figure \ref{fig:cliches} (b)). In addition, the 3 to 4-folding of many reflections reveals the presence of a complex domain structure. The pattern can be indexed within the monoclinic $C2/m$ cell proposed by Haumont \etal{}\cite{Haumont2009}. This cell has a comparatively large volume ($Z = 12$), with lattice constants $a_{\mathrm m}=17.602(1)$ \AA, $b_{\mathrm m}=7.762(1)$ \AA, $c_{\mathrm m}=5.510(1)$ \AA\ and $\beta=108.195(1)^\circ$ at 4.3 GPa. The unit cell vectors are sketched in figure \ref{fig:mailles} (a). For this cell, the pseudo-cubic lattice constants plotted in figure \ref{fig:param} were calculated as $a_{\mathrm{pc}} = a_{\mathrm m}/2\sqrt{5}$, $b_{\mathrm{pc}}=b_{\mathrm m}/2$ and $c_{\mathrm{pc}} = c_{\mathrm m}/\sqrt{2}$. At 7 GPa, a phase transition to a new high-pressure phase (phase III) is observed. The transition is evidenced by the evolution of the peak positions but is not associated with the emergence or disappearance of peaks in the diffraction pattern. The pattern could therefore be indexed within the same unit cell, but with sudden changes of the $a_{\mathrm{pc}}$ and the $c_{\mathrm{pc}}$ lattice constants respectively (figure \ref{fig:param}), while the volume remains almost constant. Finally, at 11 GPa, the crystal undergoes a transition to the $Pnma$ orthorhombic structure, with a diffraction pattern shown in figure \ref{fig:cliches} (d). The lattice vectors are recalled in figure \ref{fig:mailles} (c).

The phase sequence then reads $R3c\longrightarrow\mathrm{Phase\ II}\longrightarrow\mathrm{Phase\ III}\longrightarrow Pnma$. The transition pressures are consistent with the results by Belik \etal{} \cite{Belik2009} and Haumont \etal{} \cite{Haumont2009}, except that the more subtle transition $\mathrm{Phase\ II}\longrightarrow\mathrm{Phase\ III}$, only associated with the jumps of the $a_{\mathrm{pc}}$ and $c_{\mathrm{pc}}$ lattice parameters, was overlooked in \cite{Haumont2009}. In addition, the lattice constants in the $Pnma$ phase match very well the values reported in ref. \cite{Haumont2009} from powder x-ray diffraction carried out with hydrogen as PTM, which confirms the hydrostaticity of the stress field. The precise identification of space groups for phases II and III is beyond the scope of this paper, but we recall that Belik \etal{} have proposed two orthorhombic phases for phases II and III, also sketched in figure \ref{fig:mailles} (b). This proposition is similar to the monoclinic $C2/m$ adopted here for the calculation of lattice constants in the sense that both cells have the same volume, but differ by the presence of the monoclinic distortion. 

We now turn to the observations made in non-hydrostatic pressure conditions. The evolution of the rhomboedral angle $\alpha$ in the two experiments at hydrostatic and non-hydrostatic conditions are compared in figure \ref{fig:param} (insert). The distortion decreases more slowly in the non-hydrostatic case, reflecting the onset of the deviatoric stress even at low pressures. The transition $R3c\longrightarrow$Phase II is shifted to higher pressure by almost 1 GPa. Between 5 and 7 GPa, phase II remains stable and is identical to the phase observed under hydrostatic conditions. The transition to phase III occurs at a pressure that is very close to the 7 GPa observed in the hydrostatic case. However, at 8 GPa, the crystal exhibits a phase transition to a distinct phase, labelled III-b, that was not observed in hydrostatic conditions. The diffraction pattern is shown in figure \ref{fig:cliches} (c). This pattern could be indexed in a monoclinic cell with $a_{\mathrm m} = 9.560(2)$ \AA, $b_{\mathrm m} = 5.35(2)$ \AA, $c_{\mathrm m} = 9.547(3)$ \AA\ and $\beta\approx 109.44(3)^\circ$ at 10.2 GPa, with lattice vectors sketched in figure \ref{fig:mailles} (d). In this experiment, the short $b$ axis lies almost in the direction of the loading axis of the cell and bears the larger compressive stress. The crystal exhibits again a domain structure that makes the determination of the space group difficult, but it is interesting to note that this cell is metrically very close to the unit cells observed in other bismuth based perovskites: BiCrO$_3$ and BiMnO$_3$ \cite{Darie2010,Belik2007}. We also point out that this monoclinic cell has $a_{\mathrm m}\approx c_{\mathrm m}$ and is therefore very close to a face-centered orthorhombic cell. Finally, at 11 GPa, the crystal undergoes a transition to the orthorhombic $Pnma$ structure. It is noticeable that the crystal is then almost in a single domain state, with its $c$-axis directed along the loading axis of the cell. The lattice constants in the $Pnma$ phase are very different from the parameters obtained in hydrostatic conditions (figure \ref{fig:param}), which is in agreement with the expected larger compressive stress along the loading axis. A comparison of the relative volume $V/V_0$ under both conditions (figure \ref{fig:param}) shows that the volume measured under non-hydrostatic conditions is larger than in hydrostatic conditions. 

\section{Discussion and concluding remarks}

Because non-hydrostatic stresses are unavoidable at high enough pressures in diamond-anvil cells, the influence of deviatoric stresses on powder diffraction experiments in general has been given a lot of attention (see e.g. \cite{Funamori1997,Dubrovinsky2004,Singh2009} and references therein). Non-hydrostatic conditions are known to cause broadening, shifting, and even splitting of the diffraction peaks, even in the absence of any phase transition. In addition, grain-to-grain stresses may have an effect on the structure itself, and lead to different experimental results. For that reasons, it is recommanded to check powder data against single crystal data for the identification of phase transitions. 

In this paper, we have reported two different experiments on BFO single crystals, carried out in hydrostatic and non-hydrostatic conditions. We have shown that the non-hydrostatic stress causes a significant shift in the first transition pressure, and more importantly a modification of the transition sequence, with the emergence of a new monoclinic phase III-b. The emergence of a distinct phase is a remarkable effect and shows that a non-hydrostatic stress field is a particularly critical parameter for high-pressure experiments on BFO. It is expected to be a major concern when the PTM becomes solid. From a practical point of view, great care should be taken with the PTM and the experimental conditions in general. 
  
With this in mind, we suggest that the different PTM used in literature may be the dominating cause of the different transition pressures reported from Raman spectroscopy in ref. \cite{Haumont2006} (Argon as PTM, solid above 1.3 GPa) and infrared spectroscopy in ref. \cite{Haumont2009} (solid CsI as PTM). We note the good agreement between the results obtained in this work on single crystals and the results presented in ref. \cite{Haumont2009,Belik2009} on powder samples, all experiments being performed in very good hydrostatic conditions. On the other hand, non-hydrostatic conditions cannot explain all the discrepancies in the literature. In ref. \cite{Zhu2010,Gavriliuk2008}, the low-pressure phase transitions have not been observed although the PTM used (methanol--ethanol mixture and helium respectively) should ensure very good hydrostatic conditions in this pressure range. Other sources of disagreement, such as sample stoichiometry, have to be considered. 

\section*{Acknowledgments}

The authors are grateful for precious help from the ESRF staff, especially M. Mezouar at the ID27 Beamline for the allocation of inhouse beamtime. Support from the French National Research Agency (ANR Blanc PROPER) is acknowledged.


\end{document}